\documentstyle[12pt,aaspp,epsf]{article}

\def\refitem #1! #2! #3! #4;{\hang\noindent
    \hangindent 20pt\rm #1, \rm #2, \rm #3, \rm #4.\par}
\def\bookref{\par\noindent\hangindent 20pt}

\def\wisk#1{\ifmmode{#1}\else{$#1$}\fi}
\def\Amp     {\wisk{{\langle Q_{RMS}^2\rangle^{0.5}}}}

\def\late {650}	

\def\COBE {{\it COBE}}

\def\etal {{\it et al.}}
\def\vs   {{\it vs.}}

\def\bc {\begin{center}}
\def\ec {\end{center}}

\def\be {\begin{equation}}
\def\ee {\end{equation}}

\def\worldscimargins{
\oddsidemargin 0.2cm
\topmargin -3cm \headheight 12pt \headsep 25pt
\footheight 12pt \footskip 75pt
\textheight 26.0cm \textwidth 15.5cm
\parindent 1.0cm  \baselineskip 2.6ex
\frenchspacing
}

\worldscimargins

\begin{document}

{\Large Comparison of Spectral Index Determinations}

\pagestyle{empty}

\vspace{0.2in}

\makebox[0.5in]{}\parbox[t]{3in}
{Edward L. Wright\newline
UCLA Dept. of Astronomy\newline
Los Angeles CA 90024-1562}

\bc
ABSTRACT
\ec

\makebox[0.5in]{}\parbox[t]{5.5in}
{\small\baselineskip 12pt
\textwidth 5.5in
The index $n$ of a power law power spectrum of primordial density 
fluctuations,
$P(k) \propto k^n$, has been estimated using many different techniques.
The most precise compare the {\it COBE}\,\footnotemark[1] DMR
\footnotetext[1]{The National Aeronautics and Space Administration/Goddard
Space Flight Center (NASA/GSFC) is responsible for the design, development,
and operation of the Cosmic Background Explorer (\COBE).
Scientific guidance is provided by the \COBE\ Science Working Group.
GSFC is also responsible for the development of the analysis software
and for the production of the mission data sets.}
large angular scale $\Delta T$ to
the amplitude of the large scale structure, but these are also the most 
model-dependent.  The {\it COBE}\, DMR $\Delta T$ has 
also been compared to the
degree-scale $\Delta T$ from several experiments.  And finally, a relatively
model-independent value of $n$ can derived from the {\it COBE}\, data alone,
but the small range of angular scales covered by {\it COBE}\, limits the
precision of these methods.
}

\parindent 1.0cm
\baselineskip 14pt

\section{Introduction}

In this paper I compare several different methods for determining 
the spectral index $n$ of the power spectrum of primordial density
perturbations.  All of the determinations that use {\it COBE} data
are statistically compatible with the $n \approx 1$ predicted by the
inflationary scenario.  Because the largest scales appear as large
angular scale features on the finite solid angle of sky that is
available for viewing, the statistical uncertainties in the determination
of $n$ cannot be conquered by the usual expedient of getting more data.
A careful consideration of the statistical methods used to analyze the
large-angular scale {\it COBE} data is needed.

\section{Biased Statistics}

As an example of the pitfalls of statistical analysis, consider the
maximum likelihood method applied to determining the standard deviation
of a Gaussian from a set of $N$ independent samples drawn from the
distribution.  The likelihood function is
\be
L(\mu,\sigma) = (\sqrt{2\pi}\sigma)^{-N} \prod_{i=1}^N 
\exp[-0.5(x_i-\mu)^2/\sigma^2]
\ee
which is maximized at $\mu = N^{-1} \sum_i x_i$ and 
$\sigma^2 = N^{-1} \sum_i (x_i-\mu)^2$.
While the sample mean is an unbiased estimate of the population mean,
the variance estimate is biased by a factor of
$(N-1)/N$, illustrating the fact that
maximum likelihood estimators are only {\it asymptotically} unbiased.
Even with an unbiased estimate of $\sigma$, 
the logarithm of an unbiased estimate of $\sigma$
is not an unbiased estimate of the logarithm of $\sigma$
because noise rectification by the second derivative of the
logarithm leads to a bias of $-1/4N$.
Since the value of $N$ when estimating the power in the $\ell$'th
multipole is $N \approx (2\ell+1)\Omega_{obs}/4\pi$, and since 
$\Omega_{obs}$ is $< 8\pi/3$ due to galactic contamination,
these biases will be most signifcant for the low $\ell$'s measured
by {\it COBE}.  Thus any method for determining $n$ using {\it COBE}
data should be tested using simulated data to calibrate these biases.

\begin{figure}[t]
\plotone{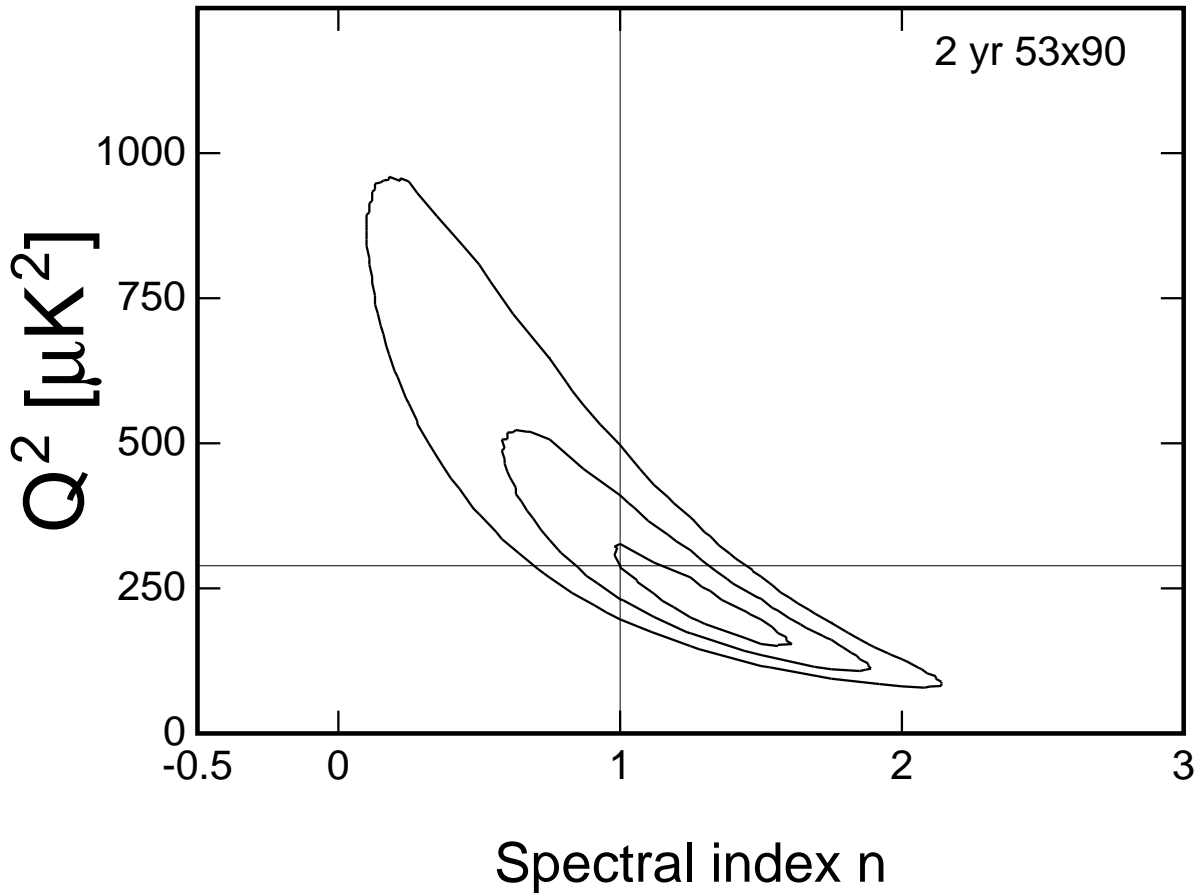}
\caption{Likelihood contours vs $Q^2$ and $n$, based on the 2 year
$53 \times 90$ cross power spectrum for $3 \leq \ell \leq 30$.}
\label{Q2vn5390}
\end{figure}

\begin{figure}[t]
\plotone{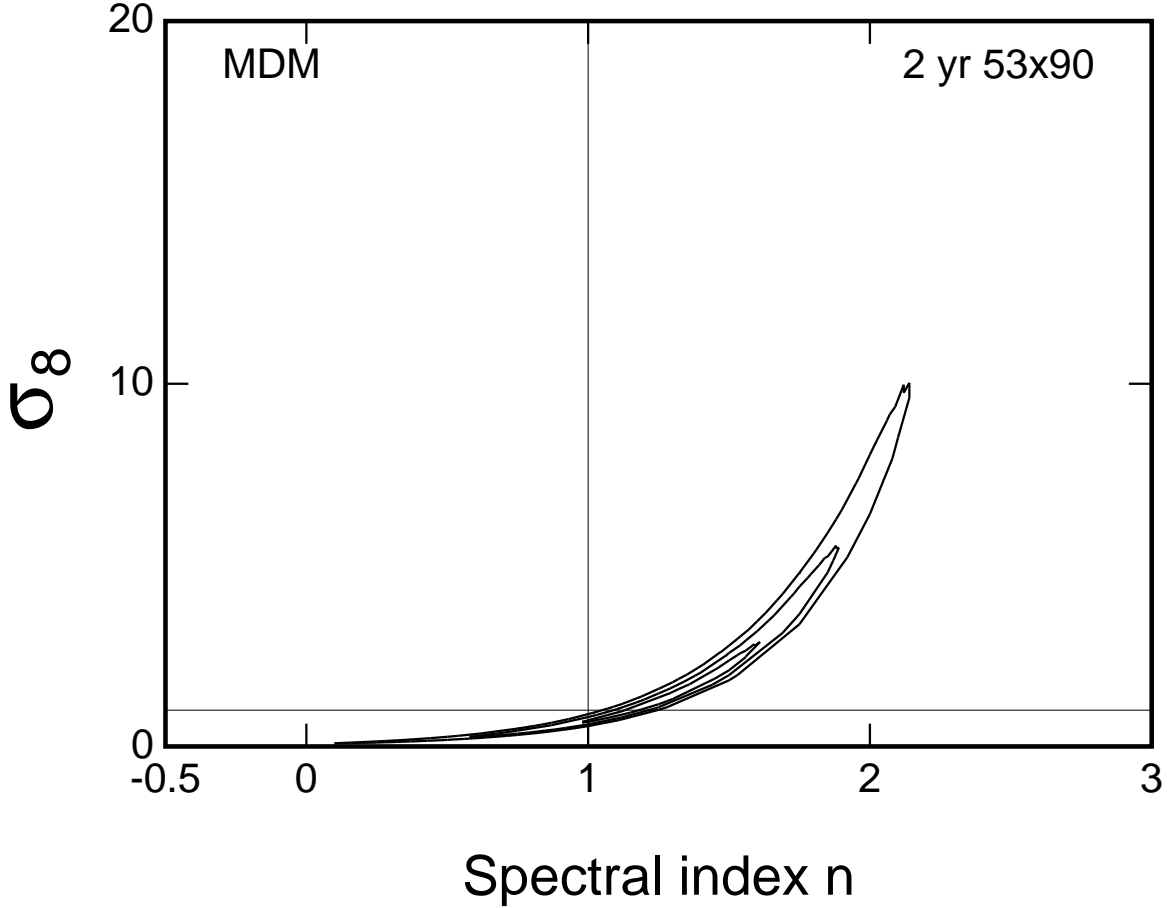}
\caption{Likelihood contours vs $\sigma_8$ and $n$, based on the 2 year
$53 \times 90$ cross power spectrum for $3 \leq \ell \leq 30$.}
\label{s8vn5390}
\end{figure}

One non-recommended technique for determining $n$ is to 
treat the integrated likelihood $f(n) = \int L(Q,n) dQ$
as a probability density for $n$.  The usual justification
for this is the Bayesian rule that the probability density
for $Q$ and $n$ after the experiment, $p_a(Q,n)$, is given by
\be
p_a(Q,n) \propto p_p(Q,n) L(Q,n)
\label{bad}
\ee
where $p_p$ is the prior density.  In this case it is true
that
\be
p_a(n) = \int p_a(Q,n) dQ \propto \int p_p(Q,n) L(Q,n) dQ.
\ee
Assuming a ``uniform'' prior to represent prior ignorance then
gives the form in Equation \ref{bad}.  But a uniform prior in $Q^2$
is not the same as a uniform prior in $\ln Q$, and they give
different values of $n$.  Different ways of expressing our
prior ignorance should not affect the answer.
A more dramatic example is shown in Figure \ref{Q2vn5390}
and Figure \ref{s8vn5390}, each showing the likelihood contours for
the Hauser-Peebles cross power spectrum of the $53 \times 90$ 
2 year maps.  In one case they are plotted versus $Q^2 = T_2^2$,
while in the other case they are plotted versus the relative mass
fluctuations in 8$h^{-1}$ Mpc spheres,
$\sigma_8 \propto T_{\approx \late}$.
The Jacobian of the transformation between $(Q^2,n)$ and $(T_{\late},n)$
depends on $n$, so a uniform prior in $(Q^2,n)$ becomes an $n$-dependent
prior in $(T_{\late},n)$.  Hence the $\int L(Q,n) dQ^2$ peaks at a much
lower $n$ than the $\int L(Q[\sigma_8,n],n) d\sigma_8$.

There is a better way, which is to use the maximum of $L(Q,n)$ for a
fixed $n$ to generate the marginal likelihood over $n$.  This approach
to ``uninteresting'' parameters is recommended by Avni (1976).  The
maximum value does not require a Jacobian when transforming to different
amplitude variables, so prior ignorance of $Q^2$ gives the same answer
as prior ignorance of $T_{\late}$.

The process of determining an amplitude parameter (usually \Amp)
and the spectral index $n$ from the {\it COBE} maps is an extreme
example of data {\it reduction}.  In this process one takes the
$360 \times 10^6$ DMR data samples per year and produces maps
with $6 \times 6144$ values, and from these maps one calculates
a smaller number of statistics.  In the final step, \Amp\ and
$n$ are estimated using the values of the statistics, leaving only
2 values derived from nearly $10^9$ input values.  This
description is general enough to describe both the G\'orski (1994)
method using linear statistics and the methods involving quadratic
statistics: the correlation function used by Bennett \etal\ (1994)
and the Hauser-Peebles power spectrum used by Wright \etal\ (1994).
The final result of any of these analysis methods is the values
$Q_{obs}$ and $n_{obs}$ determined from the real data, as well as
an estimate $\hat{\sigma_1}$ for the noise standard deviation in one
observation.

\section {Monte Carlo Simulations}

In order to test these methods for biases, it is necessary to
simulate both the {\it cosmic variance}, which gives a random map 
with random spherical harmonic amplitudes chosen from a Gaussian
distribution with a variance determined from the chosen $Q_{in}$ and
$n_{in}$, and the {\it experimental variance}, which gives the
360 million noise values needed per year.  While programs to simulate
the DMR time-ordered data do exist, none of the groups mentioned above
have worked at this level of detail.  Instead, they have used simulations
that start with the maps.

The effect of noise on the map production process can be simulated using
\be
T = \sigma_1 A^{-0.5} U
\ee
where $\sigma_1$ is the noise in one observation,
$U$ is an uncorrelated vector of unit variance zero mean Gaussian
random variables, and $A$ is the matrix with diagonal elements
$A_{ii}$ equal to the
number of times the $i^{th}$ pixel was observed, and off-diagonal elements
$-A_{ij}$ equal to the number of times the $i^{th}$ pixel was referenced
to the $j^{th}$ pixel.
Even though $A$ is singular, Wright \etal\ (1994) give a rapidly convergent
series technique for generating noise maps.  Thus each noise map
depends on 6144 independent Gaussian unit variance random variables
and the parameter $\sigma_1$.

The signal map that is added to the noise maps 
to give the ``observed'' maps is generated using independent
Gaussian random amplitudes.
Bond \& Efstathiou (1987) show that the expected variance of the coefficients
$a_{\ell m}$ in a spherical harmonic expansion of the CMBR temperature
given a power law power spectrum $P(k) \propto k^n$ is
$<a_{\ell m}^2> \; \propto 
\Gamma[\ell+(n-1)/2] / \Gamma[\ell+(5-n)/2]$
for $\ell < 40$, with the constant of proportionality chosen so that
$5<a_{2m}^2>/4\pi = Q^2$.  The simulations done by Wright \etal\ (1994)
included $\ell$'s up to 39, so the signal map depends on 1600 Gaussian
independent unit variance random variables and the two parameters
$Q$ and $n$, which I shall call $Q_{in}$ and $n_{in}$ below to distinguish
them from the fitted values.

\begin{figure}[t]
\plotone{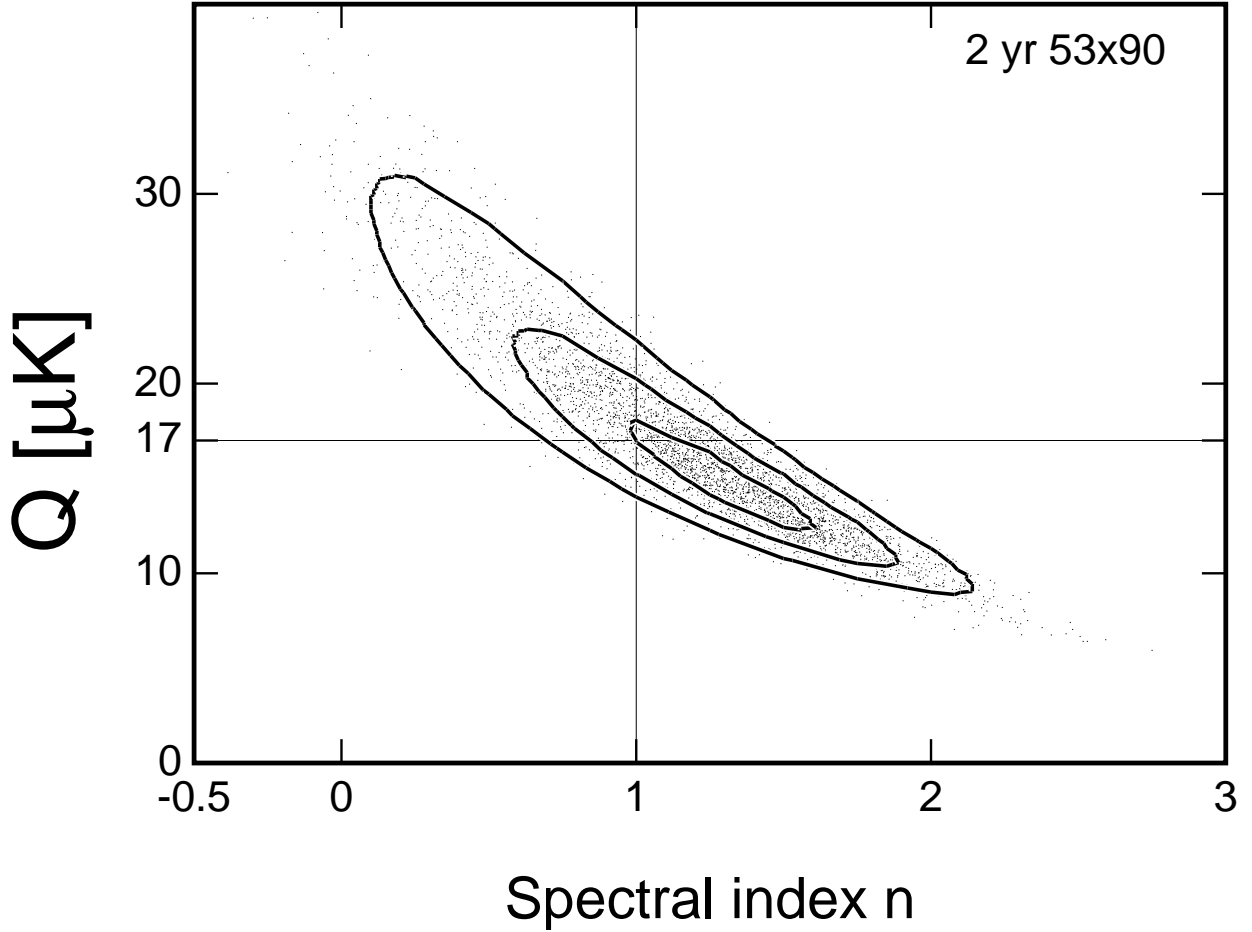}
\caption{Each point is an input parameter set that is consistent with
the real data for a given realization of the random cosmic and radiometer
variance processes.  The likelihood contours are at 
$\Delta(-2\ln L) = 1$, 4 and 9.}
\label{match}
\end{figure}

The resulting Monte Carlo map depends on a set of random variables $\{Z\}$
(1600 + 6144 elements for a one map analysis, 
or 1600 + 12288 for a cross-analysis needing two maps)
with a known distribution, and the three parameters $Q_{in}$, $n_{in}$
and $\sigma_1$.  $\sigma_1$ can be determined with great precision
using the time-ordered data.  Hence one needs to run many Monte Carlo
simulations with different values of $Q_{in}$ and $n_{in}$ and compare
the fitted values $Q_{out}$ and $n_{out}$ to the fitted values for the
real data, $Q_{obs}$ and $n_{obs}$.  For any given realization of $\{Z\}$,
the fitted values $Q_{out}$ and $n_{out}$ are a continuous function
of the input parameters $Q_{in}$ and $n_{in}$, and one can choose
values $Q_{in} = Q_{match}$ and $n_{in} = n_{match}$ 
such that $Q_{out} = Q_{obs}$
and $n_{out} = n_{obs}$.  By choosing many different realizations of
$\{Z\}$, one creates many different $Q_{match}$,$n_{match}$ pairs.
The density of the points in the $Q_{match}$,$n_{match}$ 
plane defines a probability density function
for the true parameters $Q_{true}$,$n_{true}$ that does not depend on
any prior knowledge but does depend on the experimental result in a
reasonable way.  
The random
element in the process comes from $\{Z\}$, whose properties are known.
Figure \ref{match} shows this cloud of points for the 2 year
$53 \times 90$ cross-power spectrum.  
The bias ($\Delta n = 0.1$) in the Gaussian approximation
maximum likelihood method applied to the quadratic power spectrum statistics
is only $0.25\sigma$, so the shift between the points and the contours is
hard to see.

The method using linear statistics (G\'orksi 1994) has the advantage
that the Gaussian expression for the likelihood is exact.
A further advantage of this method is that any non-singular linear
transformation of the basis functions will give the same answer,
since the covariance matrix will change to cancel the change in the values
of the statistics.
While this means that the original motivation for generating a set of
basis functions orthonormal in the cut sky is lost, one still has the
advantage that basis functions orthogonal to any number of low order
multipoles are easy to find.
G\'orski \etal\ (1994) find (for $3 \leq \ell \leq 30$) that the 
maximum of $L(Q,n)$ occurs at $n = 1.02$ for the combined
2 year 53 GHz plus 90 GHz map, and Monte Carlo simulations show that
the bias in this application of the maximum likelihood method is small.

\begin{figure}[t]
\plotone{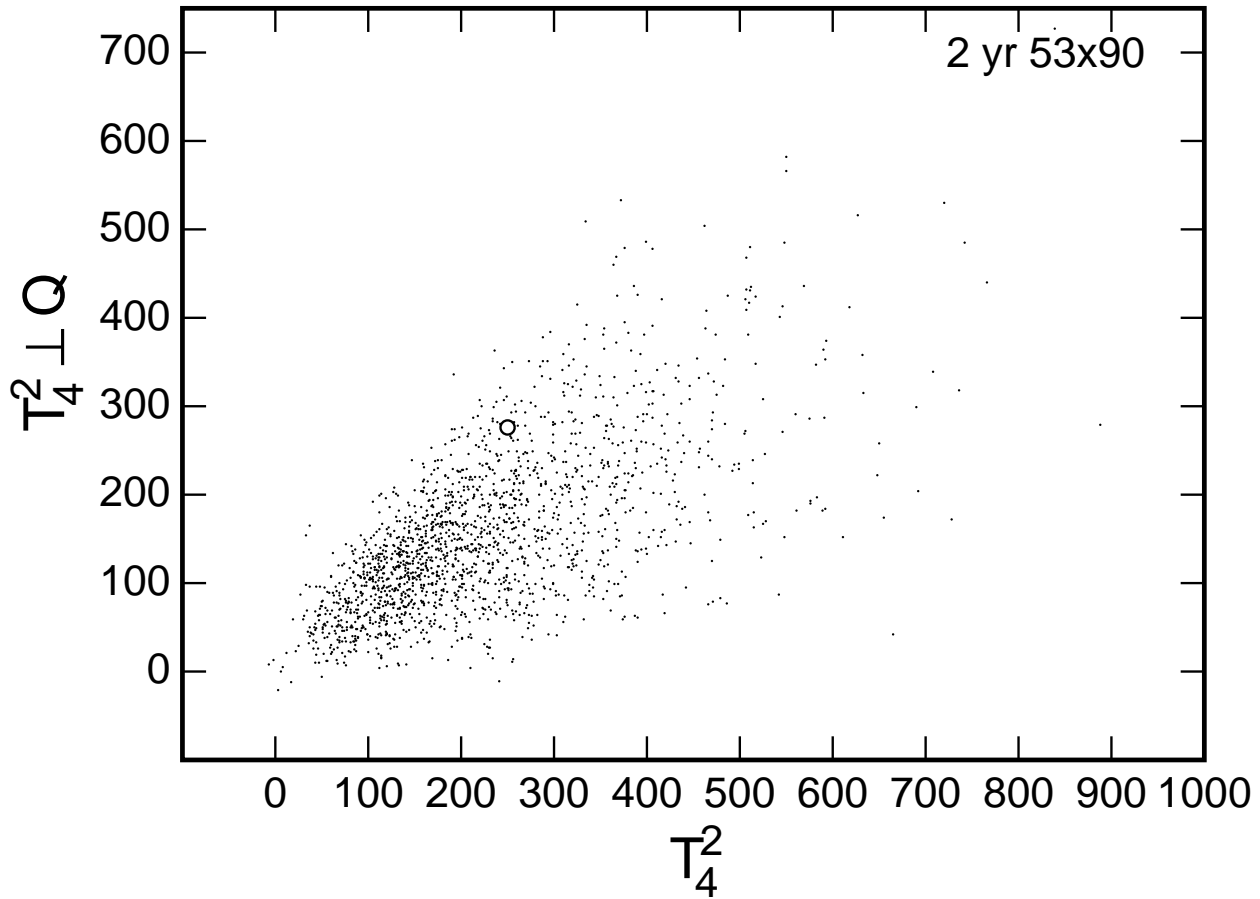}
\caption{Hexadecapole power determined using the MD method on the $x$-axis
\vs\ the MDQ method on the $y$-axis for
the real sky (open circle) and $n = 1$, \Amp\ = 17 $\mu$K
simulations.}
\label{t4}
\end{figure}

Note that the $3 \leq \ell \leq 30$ cross power spectrum fits 
in Wright \etal\ (1994) still include
the off-diagonal effect of the quadrupole on higher $\ell$'s, 
while those in G\'orski \etal\ (1994) are
completely independent of the quadrupole.
The modified Hauser-Peebles method in Wright \etal\ (1994) 
uses basis function defined using
\be
G_{\ell m} = F_{\ell m}
 - {{F_{00}<F_{00}F_{\ell m}>}\over{<F_{00}F_{00}>}}
 - \sum_{m^\prime=-1}^1 {{F_{1m^\prime}<F_{1m^\prime}F_{\ell m}>}
                              \over{<F_{1m^\prime}F_{1m^\prime}>}}.
\label{MD}
\ee
where the $F_{\ell m}$ are real spherical harmonics and the inner product
$<fg>$ is defined over the cut sphere.  These functions $G_{\ell m}$
are orthogonal to monopole and dipole terms on the cut sphere.
Call this the MD
method since the basis functions are orthogonal to the monopole and dipole.
Let the MDQ method use basis functions orthogonal to the monopole, dipole
and quadrupole:
\be
G^\prime_{\ell m} = F_{\ell m}
 - {{F_{00}<F_{00}F_{\ell m}>}\over{<F_{00}F_{00}>}}
 - \sum_{m^\prime=-1}^1 {{F_{1m^\prime}<F_{1m^\prime}F_{\ell m}>}
                              \over{<F_{1m^\prime}F_{1m^\prime}>}}
 - \sum_{m^\prime=-2}^2 {{F_{2m^\prime}<F_{2m^\prime}F_{\ell m}>}
                              \over{<F_{2m^\prime}F_{2m^\prime}>}}.
\label{MDQ}
\ee
Changing from the MD method to the MDQ method
causes the mean power in $T_4^2$ for $n = 1$ Monte Carlo skies
to go down by 31\% while $T_4^2$ for the real sky goes up by 16\%.
This leads to a higher $\ell = 4$ point and a lower value of $n$
($n = 1.02$ instead of 1.22 for the $53 \times 90$ cross-power spectrum).
Figure \ref{t4} shows the hexadecapole power in $\mu$K$^2$ measured
two different ways: on the $x$ axis the MD method; 
and on the $y$ axis $T_4^2$ measured using the MDQ method.  
The real sky is shown as the open
circle, while the dots are $n = 1$, $\Amp = 17\;\mu$K Monte Carlo
simulations.  One sees that the real sky is moderately far out
on the upper edge of the cloud of simulations, and this produces the
$0.5 \sigma$ shift in $n$ when changing basis functions.  One
also sees that the distribution of $T_4^2$ is quite skewed, which
explains the bias in the method that maximizes the
Gaussian approximation to the likelihood.

\begin{figure}[t]
\plotone{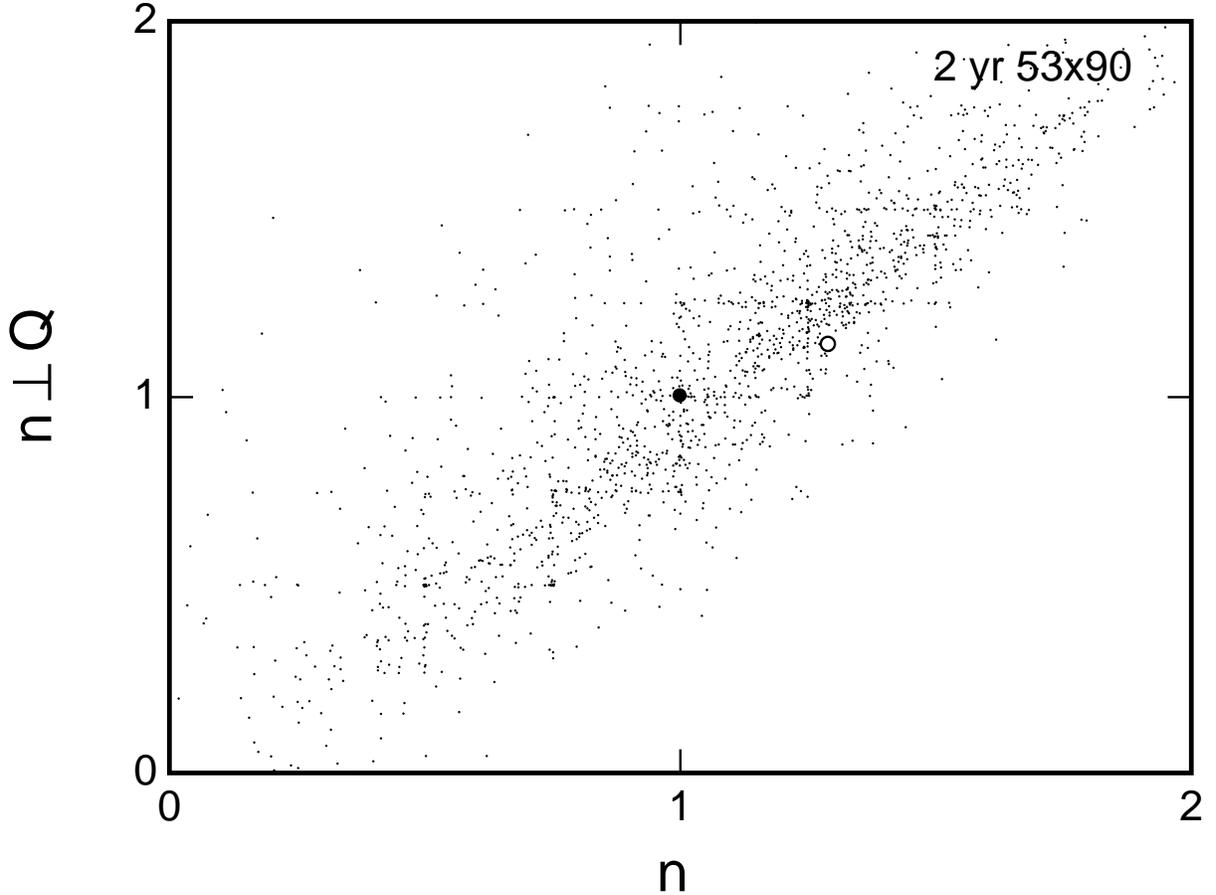}
\caption{Spectral index $n$ using the MD method on the $x$-axis
\vs\ $n$  from the MDQ method for 1800 Monte Carlo skies with
$n_{in} = 1$ and $Q_{in} = 17\;\mu$K.  The real sky is the open circle.}
\label{nvn}
\end{figure}

Figure \ref{nvn} shows the maximum likelihood values of $n$ 
from fits to $3 \leq \ell \leq 30$ for
1800 Monte Carlo runs with $n_{in} = 1$ and $Q_{in} = 17\;\mu$K.
The $x$-axis shows $n$ computed using the MD method, while the $y$-axis 
shows the results of the MDQ method.  
The real sky is shown as the open circle, and the mean of
the 1800 Monte Carlo spectra is shown as the closed circle.
This figure shows that the two methods are generally consistent,
with the real sky moderately far out in the scatter.  (The linear
features for $n = 1.00$, 1.25 and 1.50 are caused by the interpolation
among input values spaced by $\Delta n_{in} = 0.25$ during the maximum
likelihood fits.)  The overall performance of the MD method is better,
with a bias that is 20\% smaller and a standard deviation that is
8\% smaller than those given by the MDQ method.  Of course
fits that include the quadrupole ($2 \leq \ell \leq 30$)
do even better, since they use more information.
So the final result of the power spectral analysis is ambiguous:
$n \approx 1.4$ if the quadrupole is included in the fit,
$n \approx 1.25$ excluding the quadrupole using the MD method,
or $n \approx 1.0$ when rigorously
excluding the quadrupole using the MDQ method.
The existence of all these options raises the specter of ``optional
stopping'', a time-honored method of introducing systematic errors
into measurements.  But fortunately this whole range of values is
within the statistical uncertainty.

Even in very simple cases this level of disagreement between different
estimation techniques is common.  For example, the RMS difference
between the median and the mean of a set of Gaussian random numbers is
$3/4$ of the standard deviation of the mean.

\section{Degree-Scale}

The experiments at $\approx 1^\circ$ scale offer the possibility of
a better determination of the primordial power spectrum index $n$,
but the model-dependent effects of the wing of the Doppler peak
at $\ell \approx 200$ must be allowed for.  
Even in the large angle region $\ell < 30$ small model-dependent
corrections must be made.  A
Cold Dark Matter (CDM)
model with a primordial spectral index $n_{pri} = 0.96$ has an apparent 
index $n_{app}$ = 1.1 due to the ``toe'' of the Doppler peak that
extends into the $\ell < 70$ region.
Wright \etal\ (1994) have made this comparison with degree scale
experiments, and the resulting value for $n_{pri}$ is given in 
Table \ref{ntable}.

\begin{figure}[t]
\plotone{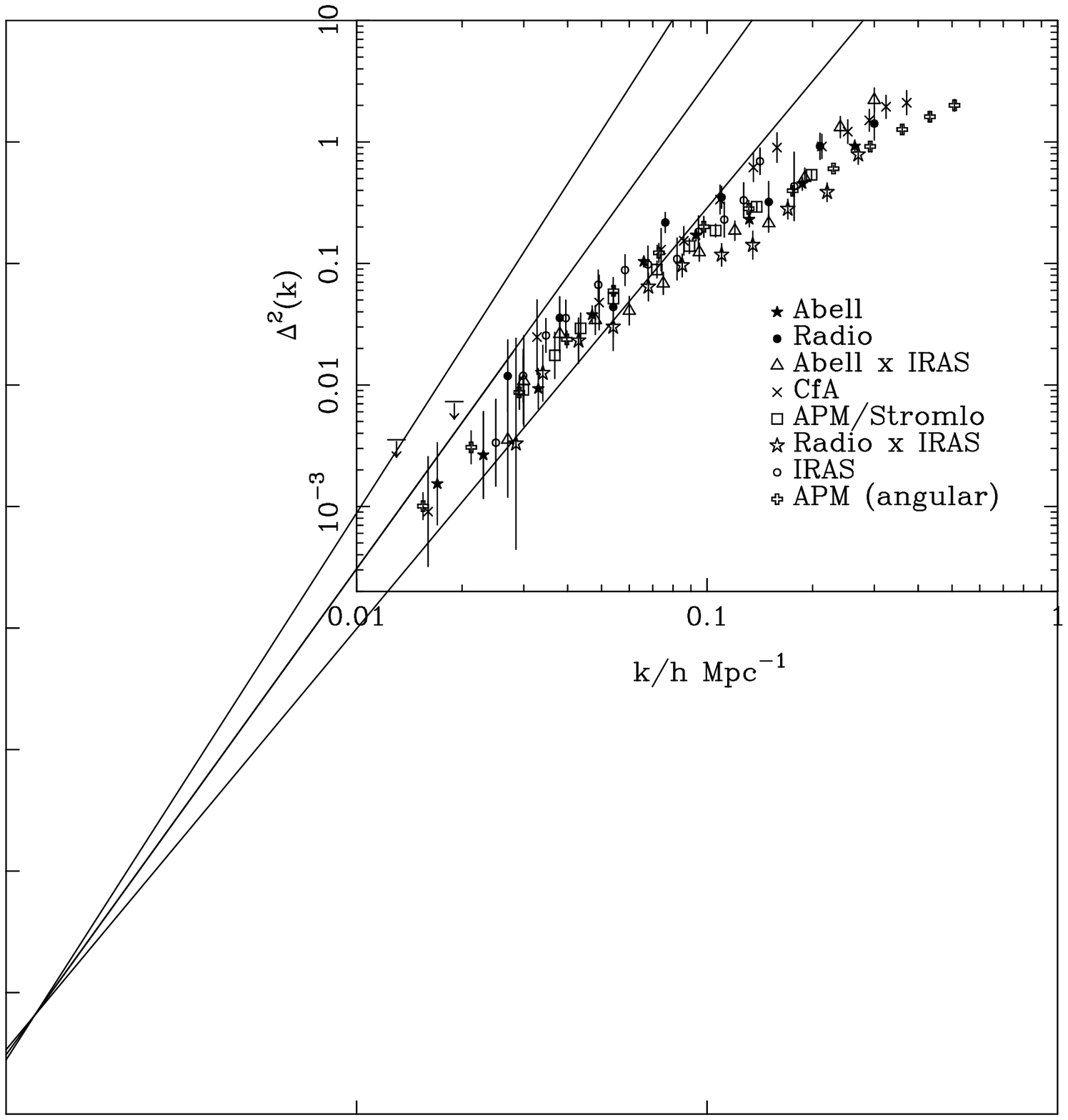}
\caption{An extended version of Figure 6 from Peacock \& Dodds, showing
$n = 0.5$, 1 and 1.5 extrapolations through the {\it COBE}\, point.}
\label{fig6}
\end{figure}

\section{Large Scale Structure}

A comparison of the extremely large scale structure seen by {\it COBE}\,
to the large scale structure seen in studies of the clustering of
galaxies also leads to an estimate of $n_{pri}$.  The uncertainty
in this method is decreased because of the large range of scales
covered, but also increased due uncertainties in the models of
large scale structure formation.  However, this comparison strongly
favors $n = 1$.  Prior to the {\it COBE}\, announcement of anisotropy,
Peacock (1991) gave a implicit prediction that for $n = 1$ the
amplitude of $\Delta T$ should be $\Amp = 18.8 \; \mu$K.
Peacock \& Dodds (1994) have extended this analysis of large scale
structure and I get a result $n_{pri} = 0.99 \pm 0.16$ from their paper
after correcting for their incorrect $\Amp = 15 \; \mu$K and increasing the
uncertainty to allow for the uncertainty in the IRAS bias, $b_I$.
In Figure \ref{fig6}
I have ``extended'' Figure 6 from Peacock \& Dodds to include
the {\it COBE}\, datum, and show extrapolations with $n = 0.5,\;1,\;\&\;1.5$
through the {\it COBE}\, point.
This result assumes that $\Omega = 1$, but Peacock \& Dodds have
also found that $\Omega^{0.6}/b_I = 1.0 \pm 0.2$.

\section{Summary}

In conclusion, both the {\it COBE}\, $\Delta T$ data alone and the ratio 
of the {\it COBE}\, $\Delta T$ data to $1^\circ$ scale $\Delta T$ are
consistent with the $n \approx 1$ prediction of the inflationary
scenario.  Furthermore, the implied level of gravitational potential
perturbations is sufficient to produce the observed large scale
(100 Mpc) structure if both the $n = 1$ and $\Omega = 1$ predictions
of inflation are correct, and the Universe is dominated by Dark Matter.

\begin{table}[t]
\begin{center}
\begin{tabular}{|llcll|}
\hline
Method & COBE dataset & Q? & Result & Reference \\
\hline
Correlation function & 1 year 53$\times$90 & N &
$n_{app} = 1.15^{+0.45}_{-0.65}$ & Smoot {\it etal} ('92) \\
COBE\,:\,$\sigma_8$      & 1 year 53+90        & N &
$n_{pri} = 1 \pm 0.23 $ & Wright {\it etal} ('92) \\
Genus \vs\ smoothing & 1 year 53           & Y &
$n_{app} = 1.7^{+1.3}_{-1.1}$ & Smoot {\it etal} ('94)  \\
RMS \vs\ smoothing   & 1 year 53           & Y &
$n_{app} = 1.7^{+0.3}_{-0.6}$ & Smoot {\it etal} ('94)  \\
Correlation function & 2 year 53$\times$90 & Y &
$n_{app} = 1.3^{+0.49}_{-0.55}$ & Bennett {\it etal} ('94) \\
Correlation function & 2 year 53$\times$90 & N &
$n_{app} = 1.1^{+0.60}_{-0.55}$ & Bennett {\it etal} ('94) \\
COBE\,:\,$1^\circ$ scale & 2 year NG           & N &
$n_{pri} = 1.15\pm 0.2$ & Wright {\it etal} ('94) \\
Cross power spectrum & 2 year 53 \& 90 & N &
$n_{app} = 1.25^{+0.40}_{-0.45} $  & Wright {\it etal} ('94) \\
Cross power spectrum & 2 year 53 \& 90 & Y &
$n_{app} = 1.39^{+0.34}_{-0.39} $  & \\
Orthonormal functions & 2 year 53+90 & N &
$n_{app} = 1.02 \pm 0.4 $ & G\'orski {\it etal} ('94) \\
COBE\,:\,$\sigma_{100}$ & 1 year 53 & Y &
$n_{pri} = 1.0 \pm 0.16 $ & Peacock \& Dodds \\
\hline
\end{tabular}
\end{center}
\caption{Spectral index determinations}
\label{ntable}
\end{table}

\section{References}

\refitem
Avni, Y. 1976! ApJ! 210! 642-646;

\refitem
Bennett, C. L., Kogut, A., Hinshaw, G., Banday, A. J., Wright, E. L., 
Gorski, K. M., Wilkinson, D. T., Weiss, R., Smoot, G. F., Meyer, S. S.,
Mather, J. C., Lubin, P., Lowenstein, K., Lineweaver, C., Keegstra, P.,
Kaita, E., Jackson, P. D. \& Cheng, E. S. 1994! ApJ! TBD! TBD;

\refitem
Bond, J. R. \& Efstathiou, G. 1987! M.N.R.A.S! 226! 655-687;

\bookref
G\'orski, K. M. 1994, submitted to the ApJL.

\bookref
G\'orski, K. M., Hinshaw, G., Banday, A. J., Bennett, C. L., Wright, E. L.,
Kogut, A., Smoot, G. F. \& Lubin, P. 1994, submitted to the ApJL.

\refitem
Hauser, M. G. \& Peebles, P. J. E. 1973! ApJ! 185! 757-785;

\refitem
Peacock, J. A. 1991! MNRAS! 253! 1p-5p;

\refitem
Peacock, J. A. \& Dodds, S. J. 1994! MNRAS! 267! 1020-1034;

\refitem
Smoot, G. F. etal. 1992! ApJL! 396! L1;

\bookref
Smoot, G. \etal\ 1994, submitted to the ApJ

\refitem
Wright, E. L., Smoot, G. F., Bennett, C. L. \& Lubin, P. M. 1994!
ApJ! TBD! TBD;

\end{document}